\documentclass[%
reprint,
superscriptaddress,
amsmath,amssymb,
aps,
prb,
]{revtex4-2}

\usepackage{graphicx}% Include figure files
\usepackage{dcolumn}% Align table columns on decimal point
\usepackage{bm}% bold math
\usepackage{gensymb}
\usepackage[mathlines]{lineno}% Enable numbering of text and display math
\usepackage{color}

\begin{document}

%\preprint{APS/123-QED}

\title{Inversion-symmetry engineering in sub-unit-cell-layered oxide thin films}% Force line breaks with \\

\author{J. Nordlander}
\email{johanna.nordlander@mat.ethz.ch}
\affiliation{Department of Materials, ETH Zurich, CH-8093 Zurich, Switzerland}

\author{M. D. Rossell}
\author{M. Campanini}
\affiliation{Electron Microscopy Center, Empa, CH-8600 D\"ubendorf, Switzerland}
\author{M. Fiebig}
\affiliation{Department of Materials, ETH Zurich, CH-8093 Zurich, Switzerland}
\author{M. Trassin}
\email{morgan.trassin@mat.ethz.ch}
\affiliation{Department of Materials, ETH Zurich, CH-8093 Zurich, Switzerland}

%\date{\today}

\begin{abstract}
Inversion symmetry breaking is a ubiquitous concept in condensed-matter science. On the one hand, it is a prerequisite for many technologically relevant effects such as piezoelectricity, photovoltaic and nonlinear optical properties and spin-transport phenomena. On the other hand, it may determine abstract properties such as the electronic topology in quantum materials. Therefore, the creation of materials where inversion symmetry can be turned on or off by design may be the ultimate route towards controlling parity-related phenomena and functionalities. Here, we engineer the symmetry of ultrathin epitaxial oxide films by sub-unit-cell growth control. We reversibly activate and deactivate inversion symmetry in the layered hexagonal manganites, h-RMnO$_3$ with R = Y, Er, Tb. While an odd number of half-unit-cell layers exhibits a breaking of inversion symmetry through its arrangement of MnO$_5$ bipyramids, an even number of such half-unit-cell layers takes on a centrosymmetric structure.  Here we control the resulting symmetry by tracking the growth \textit{in situ} via optical second-harmonic generation. We furthermore demonstrate that our symmetry engineering works independent of the choice of R and even in heterostructures mixing constituents with different R in a two-dimensional growth mode. Symmetry engineering on the atomic level thus suggests a new platform for the controlled activation and deactivation of symmetry-governed functionalities in oxide-electronic epitaxial thin films.

%\begin{description}
%\end{description}
\end{abstract}

\maketitle

%\tableofcontents

\clearpage

\section{\label{Intro}Introduction}

According to the fundamental Neumann principle, the symmetry of a material is reflected in its physical properties. Hence, whenever a symmetry is broken, new functionalities arise \cite{livio2012physics}. A special case is the presence (parity) of or absence (parity breaking) of inversion symmetry, a defining characteristic of a material that governs the emergence of parity-related functionalities. For example, in the field of quantum materials, the conservation or breaking of inversion symmetry distinguishes Dirac from Weyl semimetals \cite{Armitage2018}. Also many technologically relevant phenomena, like piezoelectricity, photovoltaics and spin-transport effects, depend on a broken inversion symmetry \cite{Trolier-McKinstry2004,Fridkin2001,Hellman2017}. 

The symmetry of a material may be broken spontaneously. For example, in ferroelectrics, spatial inversion symmetry is lost by the onset of spontaneous polarization, giving rise to the very phenomenon which established their technological relevance: piezoelectricity. However, relying on spontaneous symmetry breaking for enabling functionality in materials lacks control. It would rather be preferable to set the symmetry of a material on demand.  Recent progress in materials engineering now allows to achieve inversion-symmetry breaking by design. For example, by combining dissimilar materials to heterostructures, thus disrupting the long-range crystalline order, novel states can be created at the interface between the consituents \cite{hwang2012emergent}, resulting in phenomena like two-dimensional (2D) superconductivity \cite{ohtomo2004high} or emergent magnetic and polar properties \cite{sai2000compositional,lee2005strong,Becher2014}. In exfoliated 2D materials, the breaking of inversion symmetry on the atomic-monolayer level can lead to a unique electronic band structure \cite{Mak2010Atomically}, valley-selectivity \cite{Yao2008Valley}, electronic edge-states and nonlinear optical response \cite{kumar2013second,hsu2014second,yin2014edge}. 

All this emphasizes the fundamental importance of inversion symmetry -- or rather its absence -- for functional properties in materials and it highlights the need for deterministic control of inversion-symmetry breaking as a key aspect in state-of-the-art materials engineering. In the vast family of functional oxides, the naturally layered compounds stand out as great candidates for this purpose. In these materials, the unit cell itself is layered, and these layers exhibit a different symmetry than the unit cell in its entity. These sub-unit-cell blocks may therefore locally exhibit properties that are not permitted for the parent material. This aspect has been little explored, however.

\begin{figure*}
    \centering
    \includegraphics[scale=0.42]{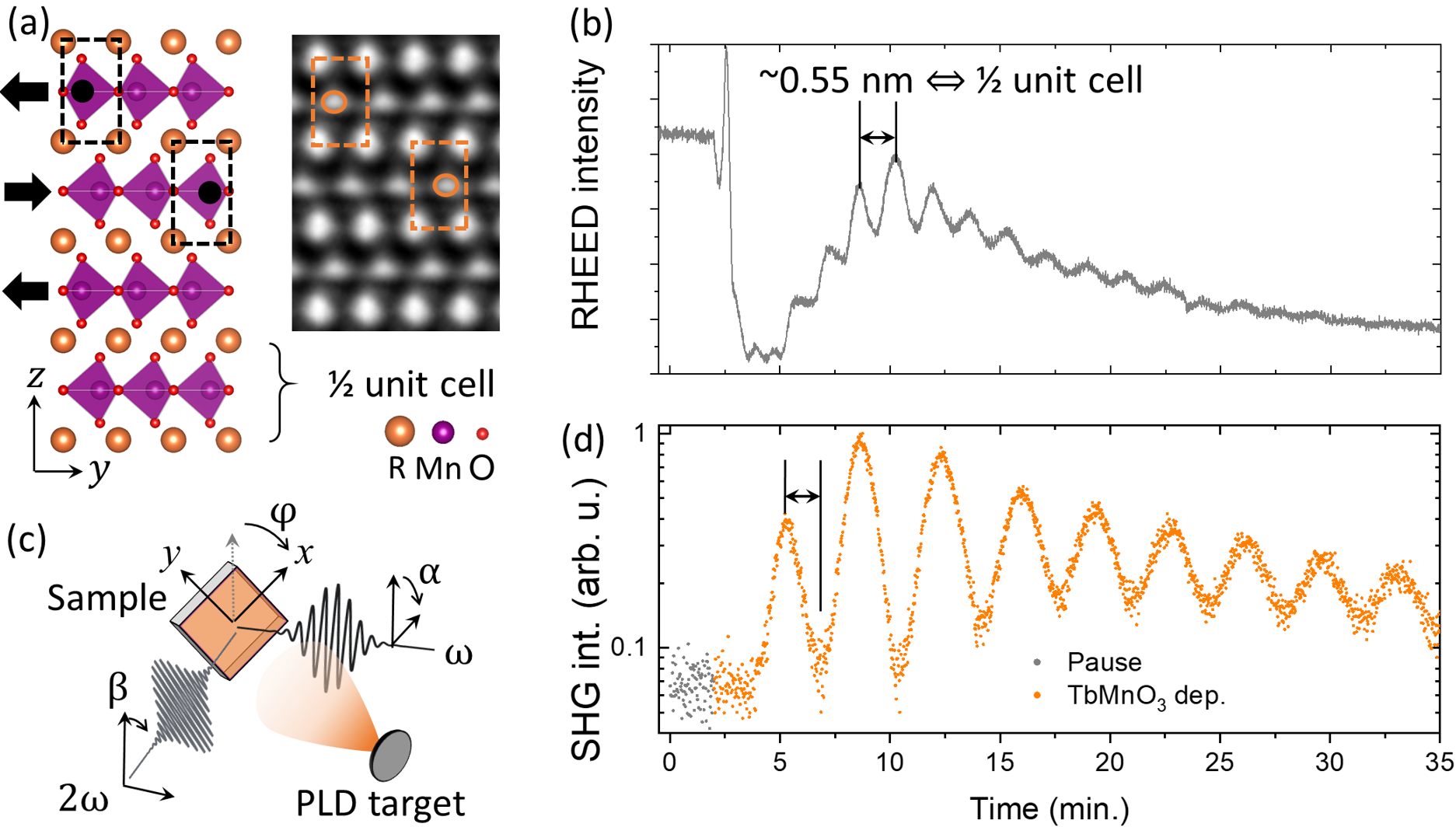} % scale=0.45 for preprint width=\columnwidth for reprint style
    \caption{(a) Prototype crystal structure of the h-RMnO$_3$ family and STEM image from a h-YMnO$_3$ film. The noncentrosymmetric structure of each half-unit-cell layer and their alternating orientation are highlighted by dashed boxes. (b) RHEED intensity oscillations indicate a layer-by-layer growth mode with each layer representing half of unit cell. (c) Experimental ISHG setup. The sample is probed in a reflection geometry in the PLD growth chamber during deposition and with simultaneous RHEED monitoring. The angle of polarization of the fundamental ($\omega$) and the SHG ($2\omega$) light are given by $\alpha$ and $\beta$, respectively. The azimuthal angle $\varphi$ denotes the orientation of the sample $x$ axis with respect to the laboratory vertical axis. (d) ISHG intensity during deposition of 10 nm (9 unit cells) of h-TbMnO$_3$ detected at ($\alpha,\beta)=(90\degree,120\degree)$. Calibration by the RHEED data reveals a periodicity of the ISHG intensity oscillations of 1 unit cell.}
    \label{fig:structure}
\end{figure*}

Here, we demonstrate symmetry engineering in ultrathin layered oxides, moving repeatedly between centrosymmetric and noncentrosymmetric states, by growth control on the sub-unit-cell level. As our model system, we choose the hexagonal manganites, h-RMnO$_3$ (R = Y, Er, Tb), because of their naturally layered structure. We deposit dielectric h-RMnO$_3$ in a layer-by-layer fashion, where each layer is only half a unit cell in height. Using in-situ optical second harmonic generation (ISHG), we probe the symmetry of the films during deposition \cite{de2017nanoscale,Nordlander2018}. We reveal that while an even number of half-unit-cell layers retain the inversion symmetry of the parent material, an odd number of these layers breaks parity because of the locally noncentrosymmetric MnO$_5$ sublattice within each half-unit-cell block. The parity-sensitive ISHG response follows this alternating symmetry in real time and allows us to set the symmetry state of the thin film system on demand, here within a thickness range of less than 6\,\AA. By expanding from a single constituent to (R$'$MnO$_3$)/(R$''$MnO$_3$) superlattices we show that this phenomenon continues across the interface between constituents, independent of R.  With our work, we thus establish layered oxides as a class of materials for exerting inversion-symmetry control and its functionalization in ultrathin epitaxial films.

\section{\label{results}Results}

The h-RMnO$_3$ thin films were grown by pulsed laser deposition on (111)-oriented yttria-stabilized zirconia (YSZ) substrates. Their crystal structure is shown in Fig. \ref{fig:structure}(a). While the h-RMnO$_3$ compounds are usually found in a noncentrosymmetric improper ferroelectric phase, a suppression of the polar mode in the ultrathin regime places the system in the paraelectric phase during the epitaxial deposition \cite{nordlander2019ultrathin}. In this phase, the unit cell is centrosymmetric and belongs to the point group 6/$mmm$. It consists, however, of two identical noncentrosymmetric half-unit-cell layers rotated by 60\degree\ with respect to each other. The symmetry of these is $\overline{6}m2$ because of the trigonal structure of the MnO$_5$ sublattice \cite{Degenhardt2001}.

For the epitaxial thin films grown by PLD, in-situ reflection high-energy electron diffraction (RHEED) intensity oscillations and post-deposition thickness analysis by x-ray reflectivity indicate a layer-by-layer growth mode where each layer corresponds to half a unit cell in height [Fig.~\ref{fig:structure}(b)]. Therefore, through precise growth control, either a centrosymmetric state (even number of half-unit-cell layers) or a noncentrosymmetric state (odd number of half-unit-cell layers) should be obtained.
 
We begin by verifying the symmetry of the half-unit-cell layers in h-TbMnO$_3$ films. To access and control the thin-film properties in real time, we use ISHG during the thin-film synthesis. This symmetry-sensitive technique allows to probe functional properties in thin films remotely and directly, as they emerge during growth. The experimental setup is sketched in Fig. \ref{fig:structure}(c). SHG is a nonlinear optical process which describes the frequency doubling of light in a material. In the electric-dipole approximation, it is described as 

\begin{equation}
    P_i(2\omega) = \epsilon_0 \chi^{(2)}_{ijk} E_j(\omega) E_k(\omega),
    \label{eq:SHG}
\end{equation}

where $E_{j,k}(\omega)$ are the electric-field components of the incident fundamental beam and $P_i(2\omega)$ denotes the resulting nonlinear polarization in the material which acts as source for the emitted SHG light. The process is parametrized by the second-order susceptibility tensor, $\chi^{(2)}$. The simultaneous monitoring of RHEED and ISHG intensity allows us to correlate the symmetry properties of the thin film with its thickness and growth mode \cite{de2017nanoscale}.

\begin{figure}
    \centering
    \includegraphics[width=\columnwidth]{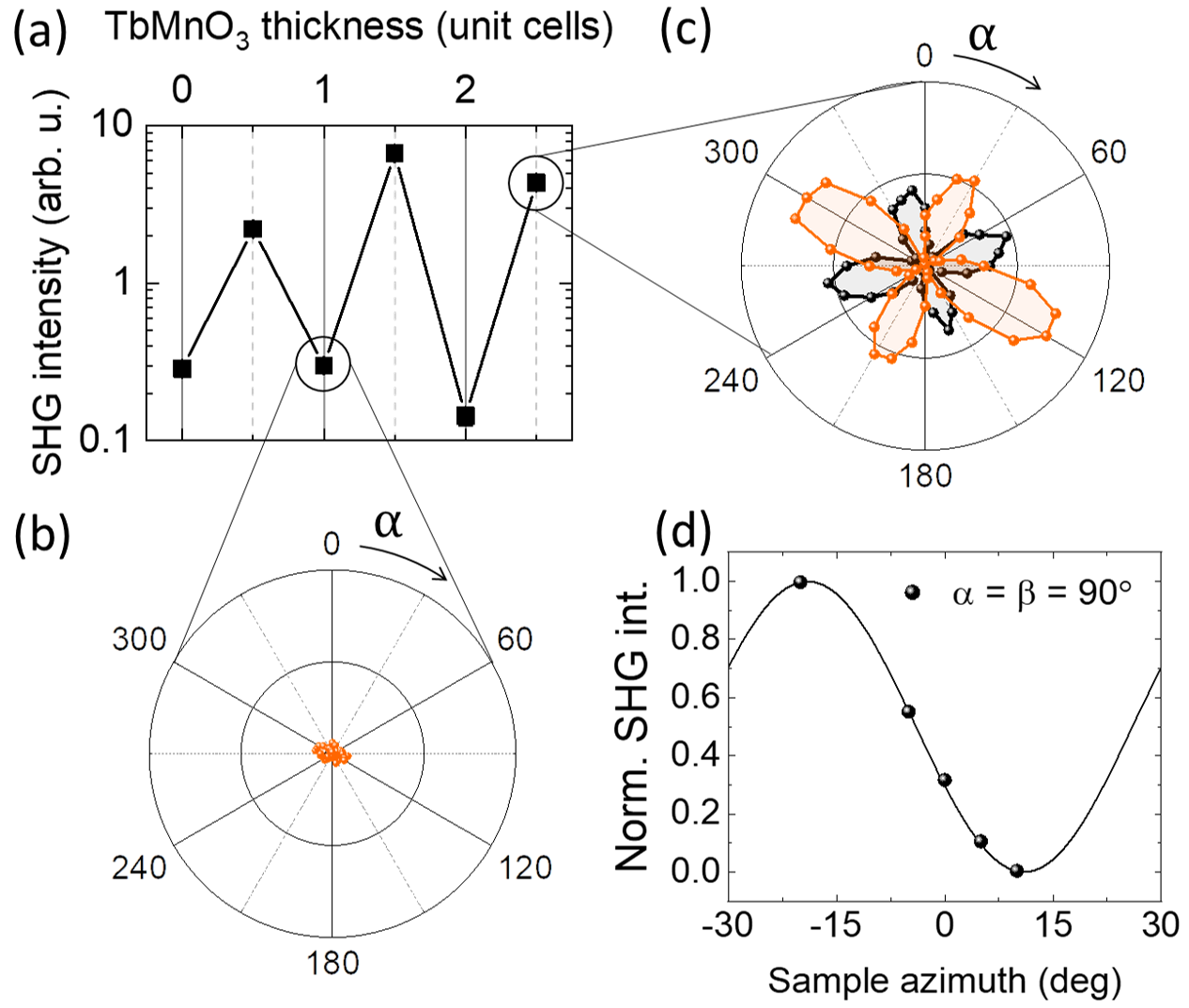} % scale=0.5 for preprint width=\columnwidth for reprint style
    \caption{(a) SHG intensity at $(\alpha,\beta)=(120\degree,90\degree)$ vs. layer thickness in a h-TbMnO$_3$ film at the growth temperature. Minima and maxima as in Fig.~\ref{fig:structure}(d) are reproduced. (b,c) Dependence of the SHG intensity for varying angle $\alpha$ and fixed angle $\beta$ at 0\degree\ (black) and 90\degree\ (orange) is shown for (b) an even and (c) an odd number of half-unit-cell layers of h-TbMnO$_3$. (b) and (c) are plotted to the same scale. Because of inversion symmetry, no SHG is detected in (b). The polarization dependence in (c) is compatible with contributions from the nonlinear susceptibility components permitted by the $\overline{6}m2$ point symmetry of the half unit cell. We attribute the slight asymmetry between the four lobes in (c) to the azimuthally varying reflectivity with the 90\degree\ reflection geometry of the ISHG setup. (d) Dependence of  half-unit-cell SHG intensity at $\alpha = \beta = 90\degree$ on the azimuthal orientation $\varphi$ of the sample. The 60\degree\ periodicity of the data further supports the $\overline{6}m2$ point group}
    \label{fig:half_uc_SHG}
\end{figure}

The real-time evolution of the ISHG signal while half-unit-cell layers are added one-by-one during the deposition of h-TbMnO$_3$ on YSZ is shown in Fig. \ref{fig:structure}(d). A periodic modulation of the ISHG signal is observed where the intensity oscillates with a period of one unit cell. For an even number of half-unit-cell layers, no SHG is detected. In contrast, an odd number of half-unit-cell layers results in a sizeable SHG intensity. Strikingly, the RHEED signal oscillates twice as fast as the SHG signal and therefore reveals the same surface morphology in both cases. This directly excludes surface-morphology-related effects, such as a step density variation during the layer-by-layer growth \cite{Jahnke1999}, as possible origin of the ISHG modulation. Instead, we attribute this modulation to the alternating symmetry of the film that occurs with the deposition of each half-unit-cell layer, as described above.

To verify this hypothesis, we analyze the polarization properties of the SHG signal (Fig.~\ref{fig:half_uc_SHG}). The the SHG response for an even and an odd number of half-unit-cell layers are shown in Figs. \ref{fig:half_uc_SHG}(b) and (c), respectively. The four-lobed symmetry seen for 2.5 unit cells in Fig. \ref{fig:half_uc_SHG}(c) is compatible with the proposed $\overline{6}m2$ point group of a half-unit-cell layer, where the allowed $\chi^{(2)}$ components in Eq.~\ref{eq:SHG} are \cite{birss1964symmetry}: $\chi_{yyy} = - \chi_{yxx} = - \chi_{xxy} = - \chi_{xyx}$, with $x$ lying parallel to the crystallographic $a$ axis. As seen in Fig.~\ref{fig:half_uc_SHG}(d)), the relation of the SHG signal to the trigonal half-unit-cell lattice is further supported by the 60\degree\ periodicity the SHG intensity with respect to rotation of sample around its $z$-axis. Here, the SHG source term for the point group $\overline{6}m2$ dictates $P^{(2\omega)} \propto \cos(3\phi)\chi^{(2)}$.

Given the polarization-independent absence of SHG for even numbers of half-unit-cell layers [Fig.~\ref{fig:half_uc_SHG}(b)], we conclude, that the ISHG intensity oscillations seen in Fig.~\ref{fig:structure}(d) are due to destructive interference of identical antiphase SHG waves from two adjacent half-unit-cell layers, where the antiphase relation comes from their relative 60\degree\ rotation, yielding a prefactor $\cos(3\cdot60\degree)=-1$ between the respective SHG source terms. Macroscopically, this destructive interference is in line with the vanishing $\chi^{(2)}$ tensor for the non-polar 6/$mmm$ point group of the full unit cell. Following the ISHG response during growth hence follows the alternating breaking and restoration of inversion symmetry with the deposition of each additional half-unit-cell layer.

\begin{figure}
    \centering
    \includegraphics[width=\columnwidth]{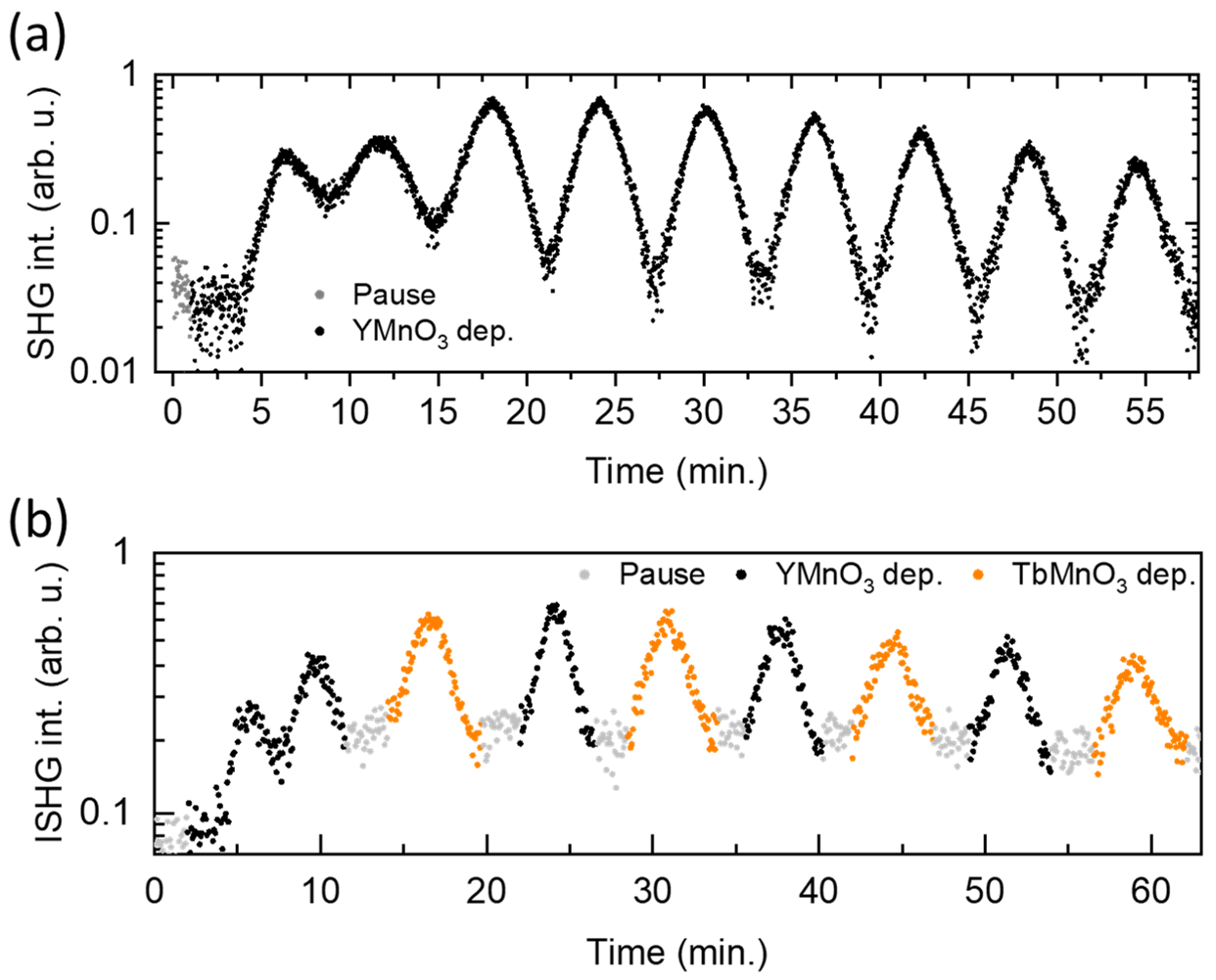} % scale=0.5 for preprint width=\columnwidth for reprint style
    \caption{ISHG intensity during h-RMnO$_3$ deposition of (a) h-YMnO$_3$ and (b) a (h-YMnO$_3$)$_1$/(h-TbMnO$_3$)$_1$ superlattice. In both cases, the measurements correspond to 10 nm (9 unit cells) of thin-film deposition and the ISHG intensity is modulated with a period of one unit cell.}
    \label{fig:oscillations}
\end{figure}

So far, we have restricted our discussion to h-RMnO$_3$ films with R = Tb. In order to determine the influence of R, we now expand our investigations towards other h-RMnO$_3$ compounds. Figure~\ref{fig:oscillations}(a) shows that the ISHG oscillations are observed also for the half-unit-cell by half-unit-cell deposition of h-YMnO$_3$ thin films. But even when we combine h-YMnO$_3$ and h-TbMnO$_3$ into a (h-YMnO$_3$)$_1$/(h-TbMnO$_3$)$_1$ superlattice, we find that the the ISHG intensity oscillation prevails (Fig. \ref{fig:oscillations}(b)). SHG from the two compounds interferes in the same way as for the single layers and at comparable oscillation amplitudes. Hence, we can conclude that the SHG signal observed here does not originate in the R-ion layer of the compound, as the signal is clearly independent of the R-ion species. Instead, it originates from the MnO$_5$ trigonal bipyramid layers uniform to all h-RMnO$_3$ compounds.

In the comparison of the data obtained by RHEED and ISHG, it is important to note that the former probes the structural integrity of a sample surface, whereas the latter senses the symmetry of the crystal lattice. We therefore find complementary information by combining the two in-situ methods, where RHEED contributes surface-morphology information and ISHG reveals the symmetry-related functionality of a specimen in real time. In particular, in the case of crystallographic defects or growth-mode variations, there can be a discrepancy between the smoothest surface (local maximum for RHEED intensity) and a defined symmetry state (local minimum or maximum for ISHG), which would manifest as a phase shift between the RHEED and the SHG oscillations. Therefore, a surface-roughness-controlled heterostructure may not necessarily be the same as a symmetry-controlled heterostructure. On the other hand, by achieving a synchronization of both surface and symmetry variations, through a synchronization of the RHEED and ISHG oscillations, we can combine the two techniques towards the design of symmetry-controlled interfaces with minimal interface roughness.

\begin{figure}
    \centering
    \includegraphics[width=\columnwidth]{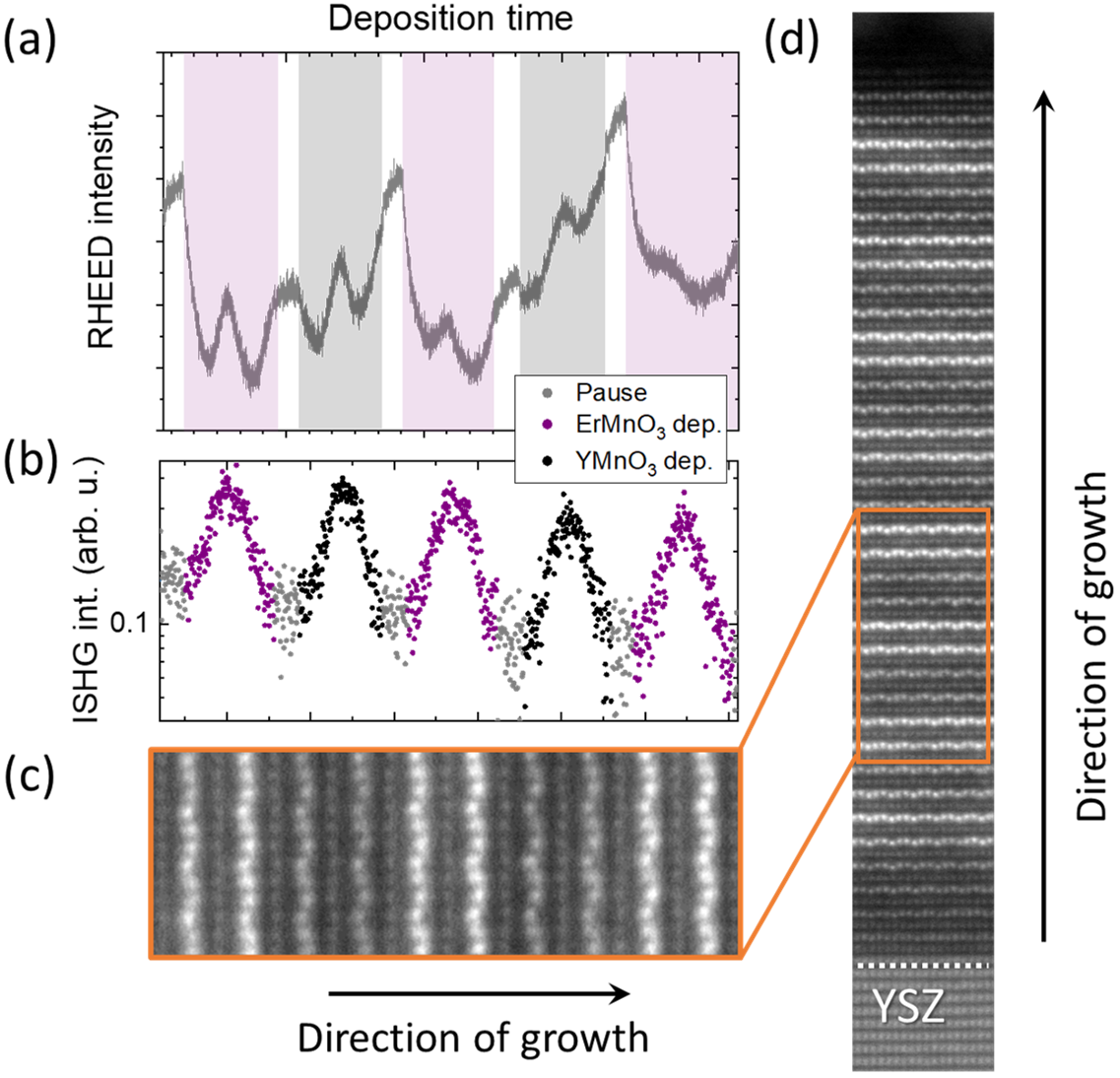} % scale=0.6 for preprint width=\columnwidth for reprint style
    \caption{(a,b) Simultaneously measured in-situ RHEED and SHG intensities during growth of a (h-YMnO$_3$)$_1$/(h-ErMnO$_3$)$_1$ superlattice. (c,d) HAADF STEM confirms atomically sharp interfaces. Note that the heavier Er atoms appear brighter than Y atoms. Further note that the periodic displacement of the R-ions along the $c$-axis indicates the occurrence of ferroelectric polarization in this superlattice. Its presence or absence is another aspect that can be growth-controlled by epitaxial constraints \cite{nordlander2019ultrathin}.}
    \label{fig:superlattice}
\end{figure}

To demonstrate this, we combined hexagonal YMnO$_3$ and ErMnO$_3$ by growing a (h-YMnO$_3$)$_1$/(h-ErMnO$_3$)$_1$ superlattice. The choice of pairing YMnO$_3$ with ErMnO$_3$ is motivated by their excellent mutual lattice matching. This increases our chances to maintain a layer-by-layer growth mode with smooth interfaces during the deposition of the heterostructure. Figure \ref{fig:superlattice} shows the simultaneous monitoring of RHEED and ISHG during growth of the (h-YMnO$_3$)$_1$/(h-ErMnO$_3$)$_1$ superlattice, where we see that the respective signal oscillations are in phase. At the completion of each layer, i.e., at each RHEED maximum, we find either a maximum or minimum in the ISHG signal. This indicates that the point during growth corresponding to a completely centrosymmetric state, also corresponds to the one with the flattest surface. We verify the expected high quality of this symmetry-controlled interface at the atomic scale using high-angle annular dark-field scanning transmission electron microscopy (HAADF-STEM). Due to the difference in atomic number for Y and Er, the two materials can be clearly distinguished in the images. This reveals perfect alternation of Y and Er layers in the heterostructure. Thus, we not only verify the coveted two-dimensional growth mode but we also confirm that no intermixing at the atomic scale occurs. The synchronization of both surface and symmetry state during thin-film deposition thereby enables the design of heterostructures with sharp interfaces and well-defined symmetry properties, which in turn opens up for deterministic control of functionality at the atomic scale.

\section{\label{end}Conclusion}

In conclusion, we have demonstrated the use of a sub-unit-cell growth mode in layered oxides for deterministic control of the resulting symmetry. Using the hexagonal manganites, h-RMnO$_3$, as model system, we select a centrosymmetric or noncentrosymmetric state of the system within the deposition of only a half-unit-cell layer. This control is enabled by the inherent noncentrosymmetry of the individual half-unit-cell layers, such that an odd number of these breaks inversion symmetry, while an even number preserves it. We have further shown that this symmetry alternation prevails in superlattices composed of different h-RMnO$_3$ and is independent of our choice of R-ions. We emphasize that the emergence of symmetry-breaking functionalities at the sub-unit-cell level is not at all limited to the h-RMnO$_3$ system. In fact, we expect similar properties in any layered material grown by sub-unit-cell layers with reduced local symmetry. In expanding beyond the aspect of inversion symmetry, we also point out that this sub-unit-cell control can be used to alternate the presence and absence of other parity-like properties like chirality, magnetic reciprocity etc. Thus, tracking and controlling thin-film oxide growth on the sub-unit-cell level has the potential to open up a new route for tailoring symmetry and coercing novel functionality in the ultrathin regime.

%%%%%%%%% END of manuscript

\bibliography{biblio}% Produces the bibliography via BibTeX.

\end{document}